\documentstyle[prd,aps,preprint,tighten]{revtex}
\begin{document}
\title{Comment on Octet Baryon Magnetic Moments in the Chiral Quark Model with
Configuration Mixing} \author{Jerrold Franklin\footnote{Internet address:
V5030E@VM.TEMPLE.EDU}} \address{Department of Physics\\
Temple University        Philadelphia, PA 19122-6082}
\date{July 17, 1998}
\maketitle
\begin{abstract}
The importance of exchange currents, and of conserving isotopic spin at both
the quark and baryon levels in application of the chiral quark model to any
calculation of baryon magnetic moments is emphasized.
\end{abstract}
\pacs{PACS number(s): 13.40.Em, 14.20.-c, 12.39.-x, 12.39.Fe}

The original static quark model made predictions for baryon magnetic
moments\cite{mtrf} that were in remarkable qualitative agreement with early
magnetic moment measurements.  However, more accurate measurements of the
magnetic moments of the baryon octet differ from the static quark model
predictions by up to  0.2 nuclear magnetons.
This difference has generally been attributed to various ``non-static" effects
in the quark model.
Some time ago, the sum rule\cite{jf69}
\begin{equation}
   ›mu(p) -›mu(n) +›mu(›Sigma^-) -
   ›mu(›Sigma^+) + ›mu(›Xi^0) - ›mu(›Xi^-)=0 ›quad  (0.49›pm .05),
\label{eq:srule}
\end{equation}
was derived, with the prediction that this linear combination of octet baryon
moments should vanish, even in the presence of a general class of non-static
effects.
 For this combination of baryons, the non-static magnetic moment contributions
would cancel out if the ultimate contribution from each quark were independent
of which baryon the quark was in.  This ``baryon independence" would follow,
for instance, if the non-static parts of the baryon wave functions were SU(3)
symmetric.  Because of the cancellation of the non-static contributions, it
was expected that the sum rule of  Eq.\ (\ref{eq:srule}) would be in better
agreement with experiment than individual quark moments.  However, subsequent
tests of the sum rule\cite{jfk} showed that it disagreed with experiment by
more than did any single magnetic moment.  The most recent experimental
value\cite{pdg} for the sum rule is shown in parentheses in Eq.\
(\ref{eq:srule}).   This violation of the sum rule indicates strong SU(3)
breaking and baryon dependent non-static contributions to the baryon magnetic
moments.

A recent paper ``Octet Baryon Magnetic Moments in the Chiral Quark Model with
Configuration Mixing"\cite{los}  has presented  the conclusion that the
sum-rule of Eq.\ (\ref{eq:srule}) is not broken by arbitrary SU(3) symmetry
breaking in the chiral quark model.   However,  the application of the chiral
quark model in Ref.›cite{los} leaves out important exchange effects
that are as large as the effects considered in Ref.   cite{los}.
These exchange effects must enter in any model if conservation of
isotopic spin is imposed at both the quark and the baryon level.
Proper inclusion of  exchange effects would produce a non zero contribution to
the sum rule of Eq.\ (\ref{eq:srule}).  The conclusion in Ref.\ \cite{los} is
also contradicted by an explicit calculation \cite{jfpi} of SU(3) symmetry
breaking within a class of models that includes the chiral quark model.

An important omission in Ref.›cite{los} is the lack of conservation of
isotopic  spin in its effective baryon wave functions, and the consequent lack
of exchange currents.   The assumption is made in Ref.     cite{los} that  the
interaction of the GBs (Goldstone bosons) is weak enough to be treated by
perturbation theory.  This assumption is used to write down effective quark
wave functions  in first order perturbation theory in  Eqs.\ (2) of Ref.\
\cite{los}.
However, because of the degeneracy of the baryons within each isomultiplet,
degenerate perturbation theory must be used.  The appropriate I-spin linear
combinations of both baryon-GB and quark-GB  must be used
in the perturbation expansion of the baryon wave functions.  Imposing isotopic
spin at the baryon level requires the inclusion of pion exchange currents,
which are absent in Ref.\ \cite{los}.

Exchange currents would show up as  the important process of a Goldstone boson
being emitted by one quark, and reabsorbed by a different quark in the same
baryon.  This emission and reabsorption would be equivalent to exchange
currents, which should contribute to the baryon magnetic moments.  Without
this process, the Goldstone boson emission considered in Ref.\ \cite{los} can
only affect the effective anomalous magnetic moments of the quarks.  Because
the reabsorption of the Goldstone boson by a different quark is left out, the
quark ``transition probabilities" listed in
Eqs.\ (A3-A5) of the appendix of Ref.\ \cite{los} do not depend on which
baryon the quark is in.
The transition probabilities in Ref.\ \cite{los}  are baryon independent in the
sense of
Ref.\  \cite{jf69},  and therefor the magnetic moments calculated in Ref.\
\cite{los} satisfy the sum rule of Eq.\ (\ref{eq:srule}).

With the inclusion of SU(3) breaking pion exchange currents as described
above, the quark transition probability for the u quark in a proton would be
different than that for a u quark in a $\Sigma^+$.  That is because a u quark
in a proton can emit a $\pi^+$ that is reabsorbed by a d quark in same proton.
There is no mechanism for this to happen in a $\Sigma^+$ hyperon where there
is no d quark to reabsorb the $\pi^+$ .  Similarly, there would be a
difference in the quark
transition probabilities between the d quark in the proton and the d quark in
the $\Xi^-$.
With SU(3) breaking (in the form of  pion dominance), the quark transition
probabilities would not be baryon independent, and the resulting baryon
magnetic moments would break the sum rule of Eq.\ (\ref{eq:srule}).
Independently of how the quarks emit or reabsorb Goldstone bosons, pion
exchange currents are
required if isotopic spin is conserved at both the quark and baryon levels
\cite{jfpi}.  This means that baryon dependent quark contributions to baryon
magnetic moments are required in any theory with full conservation of isotopic
spin, in the absence of  SU(3) symmetry.  If  SU(3) symmetry were preserved in
the GB emission, then kaon and $\eta$ exchange currents would compensate for
the pion exchange currents, preserving the sum rule.

An explicit example where the sum rule of Eq.› (1) does not
hold in a model that breaks SU(3) symmetry is given in Ref.\ \cite{jfpi}.
There, the pions, because of their anomalously light mass, are taken to dominate
the meson exchange currents, while k meson currents are left out.   The pion
exchange currents break SU(3), and the resulting baryon moments do not satisfy
the sum rule of Eq.\ (\ref{eq:srule}).   For the quark model with pion
contributions
(including exchange), the prediction from the ``QM+pion" column of Table I of
Ref.\ \cite{jfpi} for the sum rule of
Eq.\ (\ref{eq:srule}) is 0.39, which is close to the experimental value.

The model in Ref.›cite{jfpi} is quite general in that the baryon wave
functions do not depend on a specific boson emission mechanism.  The wave
functions are constrained only by conservation of isotopic spin, angular
momentum, and parity on both the quark and on the baryon levels.  Any model,
including chiral perturbation theory, that satisfied these symmetries at both
levels would be consistent with Ref.\ \cite{jfpi}.   In particular, chiral
perturbation theory (with k and $\eta$ mesons left out) should agree with the
eight equations (A1-A8)  for baryon magnetic moments listed in the Appendix of
Ref.\ \cite{jfpi}.  Those eight baryon moments are given in terms of five
parameters (bare d quark moment, bare s quark moment, pion probability in the
nucleon, effective pion orbital magnetic moment, relative importance of
decuplet baryons in the octet baryon wave function).  As long as the pion
probability is non-zero, the baryon moments given by the eight equations will
violate the sum rule in Eq.\ (\ref{eq:srule}).  A more detailed theory could
predict values for
the parameters, but could not change the parametrization without violating
isotopic spin or angular momentum conservation.

The authors of Ref.›cite{los} do conclude that configuration mixing of
SU(3) symmetric gluons\cite{lipkin} or SU(3) symmetric diquarks\cite{clf} can
produce a non zero result for the sum rule of Eq. (1).  This is surprising,
because  SU(3) symmetric mechanisms should not affect the sum rule.
The reason that these mechanisms do contribute to the sum rule in Ref.\
\cite{los} is that different mixing angles are chosen for different baryons in
these cases, which
breaks the SU(3) symmetry.  The contribution of Goldstone boson emission can
also be treated in terms of configuration mixing of the GBs in the same way as
Ref.\ \cite{los} does for gluons and diquarks.  In fact, this is the method
used in Ref.\ \cite{jfpi}.  In the case of an SU(3) breaking GB admixture, the
mixing angles are {\it required} to be different for baryons of different
strangeness because of the pion exchange mechanism discussed above.  This is
another way to
see that breaking SU(3) symmetry in the emission of Goldstone bosons must break
the sum rule of Eq. (1).

\end{document}